\documentclass[longauth]{aa}
\pdfoutput=1
\usepackage[varg]{txfonts}
\usepackage{graphicx}
\usepackage{natbib}
\bibpunct{(}{)}{;}{a}{}{,}

\begin{document}

\title{Polarimetry and flux distribution in the debris disk around HD 32297} 

\subtitle{} 

\author{R. Asensio-Torres\inst{1}
  \and M. Janson\inst{1} 
  \and J. Hashimoto\inst{2}
  \and C. Thalmann\inst{3}
  \and T. Currie\inst{4}
  \and E. Buenzli\inst{3}
  \and T. Kudo\inst{4}
    \and M. Kuzuhara\inst{5}
  \and N. Kusakabe\inst{2}
  \and L. Abe\inst{6}
  \and E. Akiyama\inst{7}
  \and W. Brandner\inst{8}
  \and T. D. Brandt\inst{9}
  \and J. Carson\inst{10,8}
  \and S. Egner\inst{4}
  \and M. Feldt\inst{9}
  \and M. Goto\inst{11}
  \and C. Grady\inst{12,13,14}
  \and O. Guyon\inst{4}
  \and Y. Hayano\inst{4}
  \and M. Hayashi\inst{7}
  \and S. Hayashi\inst{4}
  \and T. Henning\inst{8}
  \and K. Hodapp\inst{15}
  \and M. Ishii\inst{7}
  \and M. Iye\inst{7}
  \and R. Kandori\inst{7}
  \and G. Knapp\inst{16}
  \and J. Kwon\inst{17}
  \and T. Matsuo\inst{18}
  \and M. McElwain\inst{12}
  \and S. Mayama\inst{19,24}
  \and S. Miyama\inst{20}
  \and J. Morino\inst{7}
  \and A. Moro-Martin\inst{21,22}
  \and T. Nishimura\inst{4}
  \and T. Pyo\inst{4}
  \and E. Serabyn\inst{23}
  \and T. Suenaga\inst{7,24}
  \and H. Suto\inst{7,4}
  \and R. Suzuki\inst{7}
  \and Y. Takahashi\inst{7,17}
  \and M. Takami\inst{25}
  \and N. Takato\inst{4}
  \and H. Terada\inst{7}
  \and E. Turner\inst{16,23}
  \and M. Watanabe\inst{26}
  \and J. Wisniewski\inst{27}
  \and T. Yamada\inst{28} 
  \and H. Takami\inst{7}
  \and T. Usuda\inst{7}
  \and M. Tamura\inst{2,7,17}
  } 


\institute{Department of Astronomy, Stockholm University, AlbaNova University Center, SE-106 91 Stockholm, Sweden\\
e-mail: ruben.torres@astro.su.se / markus.janson@astro.su.se
  \and Astrobiology Center of NINS, 2-21-1, Osawa, Mitaka, Tokyo, 181-8588, Japan
  \and Swiss Federal Institute of Technology (ETH Zurich), Institute for Astronomy, Wolfgang-Pauli-Strasse 27,
CH-8093 Zurich, Switzerland 
  \and Subaru Telescope, National Astronomical Observatory of Japan, 650 North A'ohoku Place, Hilo, HI 96720,
USA
\and Department of Earth and Planetary Sciences, Tokyo Institute of Technology, 2-12-1 Ookayama, Meguroku,
Tokyo 152-8551, Japan
   \and Laboratoire Lagrange (UMR 7293), Universite de Nice-Sophia Antipolis, CNRS, Observatoire de la Cote
d'Azur, 28 Avenue Valrose, 06108 Nice Cedex 2, France
   \and National Astronomical Observatory of Japan, 2-21-1, Osawa, Mitaka, Tokyo 181-8588, Japan
   \and Max Planck Institute for Astronomy, K{\"o}nigstuhl 17, 69117 Heidelberg, Germany
   \and Astrophysics Department, Institute for Advanced Study, Princeton, NJ 08540, USA
   \and Department of Physics and Astronomy, College of Charleston, 68 George St., Charleston, SC 29424, USA
   \and Universitats-Sternwarte Munchen, Ludwig-Maximilians-Universitat, Scheinerstr. 1, D-81679 Munchen,
Germany
	\and Exoplanets and Stellar Astrophysics Laboratory, Code 667, Goddard Space Flight Center, Greenbelt,
MD 20771, USA
	\and Eureka Scientific, 2452 Delmer, Suite 100, Oakland, CA 96002, USA
	\and Goddard Center for Astrobiology, Goddard Space Flight Center, Greenbelt, MD 20771, USA
	\and Institute for Astronomy, University of Hawaii, 640 N. A'ohoku Place, Hilo, HI 96720, USA
	\and Department of Astrophysical Science, Princeton University, Peyton Hall, Ivy Lane, Princeton, NJ 08544,
USA
	\and Department of Astronomy, The University of Tokyo, 7-3-1 Hongo, Bunkyo-ku, Tokyo 113-0033, Japan
	\and Department of Astronomy, Kyoto University, Kitashirakawa-Oiwake-cho, Sakyo-ku, Kyoto, Kyoto
606-8502, Japan
	\and The Center for the Promotion of Integrated Sciences, The Graduate University for Advanced Studies (SOKENDAI), Shonan International Village, Hayama-cho, Miura-gun, Kanagawa 240-0193, Japan
	\and Hiroshima University, 1-3-2 Kagamiyama, Higashihiroshima, Hiroshima 739-8511, Japan
	\and Space Telescope Science Institute, 3700 San Martin Drive, Baltimore, MD 21218, USA
	\and Center for Astrophysical Sciences, Johns Hopkins University, Baltimore, MD 21218, USA
	\and Kavli Institute for Physics and Mathematics of the Universe, The University of Tokyo, 5-1-5 Kashiwanoha,
Kashiwa, Chiba 277-8568, Japan
	\and Department of Astronomical Science, The Graduate University for Advanced Studies (SOKENDAI), 2-21-1
Osawa, Mitaka, Tokyo 181-8588, Japan
	\and Institute of Astronomy and Astrophysics, Academia Sinica, P.O. Box 23-141, Taipei 10617, Taiwan
	\and Department of Cosmosciences, Hokkaido University, Kita-ku, Sapporo, Hokkaido 060-0810, Japan
	\and H. L. Dodge Department of Physics \& Astronomy, University of Oklahoma, 440 W Brooks St., Norman,
OK 73019, USA
	\and Astronomical Institute, Tohoku University, Aoba-ku, Sendai, Miyagi 980-8578, Japan} 


\abstract{We present high-contrast angular differential imaging (ADI) observations of the debris disk around HD 32297  in $H$-band, as well as the first polarimetric images for this system in polarized differential imaging (PDI) mode with Subaru/HICIAO. In ADI, we detect the nearly edge-on disk at $\geq$\,5$\sigma$ levels from $\sim$0.45$\arcsec$ to $\sim$1.7$\arcsec$ (50--192\,AU) from the star and recover the spine deviation from the midplane already found in previous works. We also find for the first time imaging and surface brightness (SB) indications for the presence of a gapped structure on both sides of the disk at distances of $\sim$0.75$\arcsec$ (NE side) and $\sim$0.65$\arcsec$ (SW side). Global forward-modelling work delivers a best-fit model disk  and well-fitting parameter intervals that essentially match previous results, with high-forward scattering grains and a ring located at 110\,AU. However, this single ring model cannot account for the gapped structure seen in our SB profiles. We create simple double ring models and achieve a satisfactory fit with two rings located at 60 and 95\,AU, respectively, low-forward scattering grains and very sharp inner slopes. In polarized light we retrieve the disk extending from $\sim$0.25--1.6$\arcsec$, although the central region is quite noisy and high S/N are only found in the range $\sim$0.75--1.2$\arcsec$. The disk is polarized in the azimuthal direction, as expected, and the departure from the midplane is also clearly observed. Evidence for a gapped scenario is not found in the PDI data. We obtain a linear polarization degree of the grains that increases from $\sim$10$\%$ at 0.55$\arcsec$ to $\sim$25$\%$ at 1.6$\arcsec$. The maximum is found at scattering angles of $\sim$90$^{\circ}$, either from the main components of the disk or from dust grains blown out to larger radii. }

\maketitle 

\begin{table*}
\caption{$H$-band ADI-based studies on the disk around HD 32297. ET = Exposure Time, FOV = Field of View, FWHM = Full Width Half Maximum, IWA = Inner working angle, cADI = classical ADI, rADI = radial ADI}
\label{table1} 
\centering
\begin{tabular}{c c c c c c c}
\hline\hline
$H$-band work & Reduction&ET on target (min)& FOV rotation ($^{\circ}$) & FWHM (mas) / seeing ($\arcsec$) & IWA ($\arcsec$)\\ 
\hline 
\textbf{Boccaletti et al. 2012}& cADI, rADI, LOCI &28.5&24.8&65 / 0.97&0.5--0.6\\
\textbf{Esposito et al. 2014} &Modified LOCI&15&30.5& 45 / 1.02--1.20&0.45\\
\textbf{This paper} &PCA, LOCI&26.5&16.9&49 / 0.8&0.45--0.5\\
\hline
\end{tabular}
\end{table*}

\section{Introduction}
Debris disks are a stage in circumstellar disk development in which most of the primordial gas has already been dissipated and new dust is formed from collisions between comets, asteroids, or planetesimals \citep[e.g.][]{wyatt2008}. For this reason, these systems can reveal the environment and circumstances of planet formation around stars other than the Sun. 

Multi-wavelength observations  are necessary to reveal the distinct morphologies and compositions of the disks. Dust emission in the thermal-infrared (IR) and sub-millimeter can be identified as photometric excess in the stellar spectral energy distribution (SED), while scattered stellar light has been directly imaged in optical and near-IR bands. Observations and modelling of the full SED are usually required to constrain the dust grain composition. High-contrast direct imaging observations, on the other hand, have the ability to uncover the grain distribution and disk structures, such as warps, spiral arms, or gaps \citep[e.g.][]{janson2016, akiyama2015, boccaletti2015,muto2012,kalas2007}, and to provide indirect evidence of the presence of unseen planets \citep{dong2015, nesvold2015,quillen2006}.

 HD 32297 is a young \citep[$\leq$\,30 Myr;][]{kalas2005} and nearby \citep[112\,pc away;][]{perryman1997} A-type star with high-IR excess, L$_\mathrm{IR}$/L$_\mathrm{*}$$\geq$ 2.7$\times$10$^{-3}$, as found by the Infrared Astronomical Satellite (IRAS) data \citep{silverstone2000}. It is also one of the few debris disks where gas has been observed. \citet{redfield2007} found NaI absorption, while \citet{donaldson2013} detected a CII line at 158\,$\mu$m. The system was first resolved in scattered light in the $J$-band as a nearly edge-on disk by \citet{schneider2005} with the NICMOS camera on the Hubble Space Telescope (HST) and, since its discovery, HD 32297 has been studied in a broad range of wavelengths.

According to optical resolved images, the disk is extending in the $R$ filter to a distance of 1680\,AU from the star \citep{kalas2005}. Together with near-IR observations, the disk shows a surface brightness (SB) asymmetry, where the south-west (SW) lobe is brighter than the north-east (NE) counterpart. While NICMOS/HST $H$ and $K$ images find this asymmetry to be significant beyond 1.5$\arcsec$ \citep{debes2009}, \citet{mawet2009} in the $K$ band confirmed a brightness inequality of 0.3--0.5\,mag/arcsec$^{2}$ between 5--10$\arcsec$ radius with the Palomar Hale Telescope. The position angles of the midplanes between the lobes were also found to differ by up to 31$^{\circ}$ by \citet{kalas2005}. This large-scale warped structure and the SB asymmetry at large projected separations is thought to be due to interaction with the interstellar medium, which was confirmed by the recent STIS/HST images \citep{schneider2014}.

Ground-based observations in the mid-IR (10--20\,$\mu$m) by \citet{moerchen2007} and \citet{fitzgerald2007} on Gemini indicated a more symmetric disk and, together with \citet{mawet2009}, suggested an interior clearing or dust deficiency out to $\sim$60--85 AU. Finally, the debris disk around HD 32297 has also been resolved at millimetre wavelengths (1.3\,mm) with the Combined Array for Research in Millimeter-wave Astronomy (CARMA) \citep{maness2008}, showing an asymmetric disk with a peak millimeter emission in the SW lobe that is much brighter than indicated by mid-IR results.

More recently, ground-based angular differential imaging (ADI) studies yielded high-contrast observations with much higher signal-to-noise (S/N) images as a result of an improved static stellar PSF halo subtraction. Together with adaptive optics (AO), which corrects for rapidly varying atmospheric aberrations, these analyses achieve a high Strehl ratio to provide an enhanced spatial resolution close to the diffraction limit of the telescope. \citet{boccaletti2012} with the Very Large Telescope array (VLT) obtained VLT/NACO $H$+$K$ observations and, together with \citet{currie2012} with NIRC2/Keck $K$ images, first made use of the coronagraphic ADI technique to study the inner disk region below $\sim$250\,AU. Later, \citet{esposito2014} modelled ADI self-subtraction in $H$- and $K$-band images to recover the original SB distribution, while \citet{rodigas2014} obtained $L'$-band (3.8\,$\mu$m)  data with LMIRcan/LBTI to test whether the far-IR \citet{donaldson2013} cometary grains model proposed from SED modelling matched the 1-4\,$\mu$m disk scattered light. 

All these works find a clear small-scale deviation of the spine of the disk from the midplane. This bent morphology occurs because the inclination is not perfectly edge-on in combination with forwards scattering grains, as proven by modelling \citep{boccaletti2012,currie2012} . The SB results are more ambiguous, but in general it appears that in $K$ \citep{currie2012} and $L$ bands \citep{rodigas2014} there is a brightness asymmetry at small separations ($\leq$ 0.7$\arcsec$) as a result of a big jump in the SW profile. \citet{boccaletti2012} and \citet{esposito2014}, however, measured a predominant symmetry in $K$-band SB profiles. These authors also found the $H$-band profiles to be largely symmetrical \citep{boccaletti2012} or to have a few NE>SW asymmetries at $\sim$90 and 160--230 AU \citep{esposito2014} . Power law breaks appear to occur at $\sim$95 and $\sim$125\,AU \citep{esposito2014,currie2012,schneider2005} which, together with modelling work, has placed the planetesimal ring at around $\sim$110\,AU in $H$ and $K$ bands \citep[e.g.][]{boccaletti2012,currie2012} or closer for \citet{esposito2014}, at $\sim$95--99\,AU, also in the $H$ band.

The HD 32297 SED has also been sampled from the optical to the millimeter in different studies and instruments. \citet{fitzgerald2007} modelled the SED with photometric data from the mid-IR to the UV, \citet{maness2008} provided 1.3\,mm observations with CARMA, and \citet{currie2012} later studied a broad range of wavelengths from 0.4--1300\,$\mu$m. Despite the strong degeneracies involved, their results lean towards a disk formed by at least two dust populations in different locations with grain sizes in the sub-$\mu$m and $\mu$m  range. Dust distribution constraints play a key role in breaking these degeneracies. \citet{donaldson2013} sampled the SED with PACS and SPIRE 63--500\,$\mu$m observations on Herschel and used previous \citet{boccaletti2012} resolved imaging modelling to constrain the geometric parameters, finding that an outer ring at 110 AU and a inner warm ring located at $\geq$1.1\,AU with a grain size $\geq$2\,$\mu$m was necessary to explain the mid-IR flux. The composition was  consistent with cometary-like grains. Finally, \citet{rodigas2014} resolved the disk in the $L'$ band (3.8\,$\mu$m) and found that the proposed cometary model gave a poor fit to the SB measurements. A modified density distribution model composed of pure water ice may best match the observations. Overall, the results appear to converge towards a multi-component disk, of which one component is a ring placed around 110\,AU.

HD 32297 is thus a very well-studied system, but it also exhibits several ambiguous characteristics that are not yet fully understood, such as the likely presence of several dust ring compositions and locations. We present high-contrast, AO-assisted images of the HD 32297 debris disk to test previous results and to obtain the basis for the study of our new polarized differential imaging (PDI) observations. Our work constitutes the first polarimetry study for this system, which has proven to provide an excellent contrast between the unpolarized stellar PSF and the polarized light scattered off from dust particles \citep[e.g.][]{perrin2015}. Induced polarization can also be used to interpret the size, shape, composition, and location of the grains \citep[e.g.][]{min2012}.

\begin{figure*}
\centering
\includegraphics*[scale=0.5]{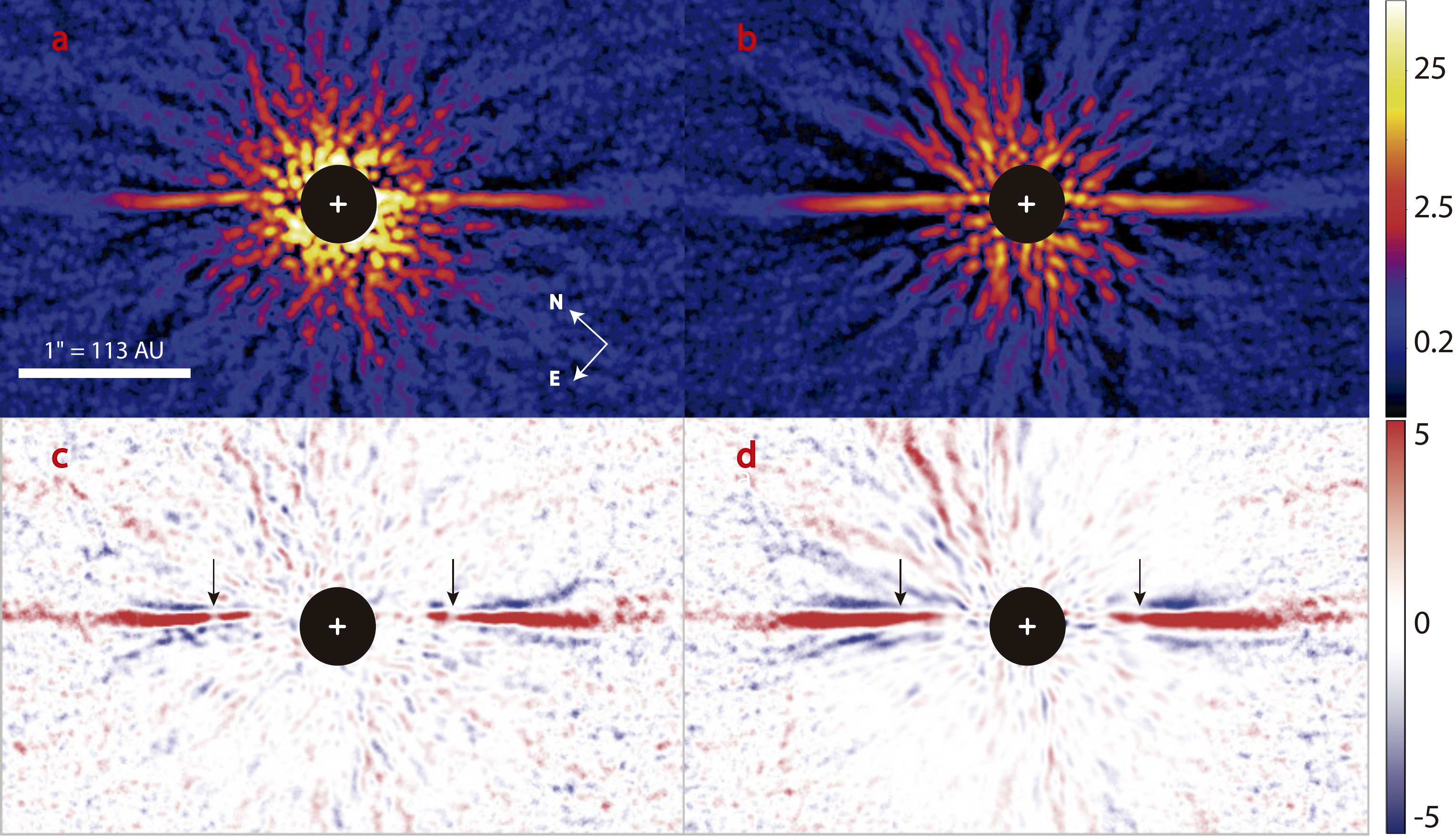}
\caption{Reduced HD 32297 ADI images. \textbf{(a)} LOCI reduction \textbf{(b)} PCA reduction, both in logarithmic stretch and units of mJy/arcsec$^{2}$. \textbf{(c)} and \textbf{(d)} correspond to \textbf{(a)} and \textbf{(b}) S/N images, respectively, in a [-5$\sigma$, 5$\sigma$] linear scale. All the images have been convolved with an aperture of diameter $\sim$FWHM. The black arrows in the S/N maps point out the locations of the potential gaps at distances of $\sim$0.75$\arcsec$ (NE side) and $\sim$0.65$\arcsec$ (SW side) from the blocked star, marked with a white cross. The diameter of the black mask is $\sim$0.4$\arcsec$.}
\label{disk}
\end{figure*}

\section{Observations}

The full intensity observations were taken on the 2015 January 12 as part of the SEEDS (Strategic Exploration of Exoplanets and Disks with Subaru/HICIAO) program at the 8.2-m Subaru Telescope located at the Mauna Kea summit, Hawaii. The IR HICIAO camera \citep{tamura2006} was used to acquire high-contrast images of the debris disk around HD 32297 in the $H$-band filter, with a central wavelength of 1.65\,$\mu$m and a bandwidth of 0.29\,$\mu$m, under a mean seeing of 0.8$\arcsec$ during the observations. The adaptive optics (AO188) capability permitted us to achieve a PSF full width half maximum (FWHM) of 5.1\,px or 48.6 mas, close to the diffraction limit, with a plate scale of 9.50 $\pm$ 0.02\,mas per pixel. Two observing blocks were taken inside the 20$\arcsec$ $\times$ 20$\arcsec$ field of view of the HICIAO camera. First, six unsaturated photometric calibration images of 1.5 seconds each were acquired, including a neutral density filter with a transmission of 0.856$\%$ in the $H$ band \citep{janson2013}. These were followed by 53 saturated science frames of 30 seconds, leading to a total exposure time of 26.5 minutes. The image rotator was controlled so as to keep the orientation of the pupil constant, allowing for ADI \citep[angular differential imaging;][]{marois2006} data reduction. Because of the limited Strehl ratio in the $H$ band, we refrained from using a coronagraph, but allowed the target star to saturate out to 12\,px = 0.11$\arcsec$ so as to get high sensitivity with low read noise and save time on overheads. A total field rotation of 16.9$^{\circ}$ was acquired between the first and last exposure of the observation. 

A different set of polarimetric imaging observations was carried out in PDI mode on 2014 October 10, also in the $H$-band filter and under the same plate scale. The rotatable half-wave plate was cycled through the retarder position angles 0$^{\circ}$, 45$^{\circ}$, 22.5$^{\circ}$, and 67.5$^{\circ}$ in which, after passing through a Wollaston prism, the incident light was split into two perpendicular linearly polarized components (I$_{\parallel}$ and I$_{\bot}$). Thus, the ordinary and extraordinary rays are obtained simultaneously, each with a 10$\arcsec$ $\times$ 20$\arcsec$ field, with polarization directions 0$^{\circ}$ and 90$^{\circ}$, 90$^{\circ}$ and 0$^{\circ}$, 45$^{\circ}$ and 135$^{\circ}$, and 135$^{\circ}$ and 45$^{\circ}$, respectively for every retarder position. A total of 20 frames of 20 seconds each were acquired in every half-wave position, which adds up to a total exposure time of 26.4 minutes. The star saturated out to a radius of 15\,px ($\sim$0.14$\arcsec$) and the use of AO provided a FWHM of 60\,mas under a seeing of 0.7$\arcsec$.

\section{Data reduction}

\subsection{Angular differential imaging}
Before flatfield and dark correction, the striped pattern (correlated read-noise) of the raw HICIAO ADI images was removed and the frames were corrected for field distortion. Absolute centring was based on visual inspection, while relative centroiding was carried out using PSF fitting on non-saturated parts of the PSF. 

We attempt to remove the HD 32297 diffracted starlight to improve the high-contrast sensitivity of the faint debris disk in close separation from the star. The ADI technique complemented by principal component analysis (PCA) is then used, which reconstructs the PSF of the central star from a library of reference PSFs and subtracts it from the target image \citep{soummer2012}. This is the so-called PCA-ADI technique that permits forwards modelling of astronomical sources, such as circumstellar disks, which in some cases has proven better at removing certain systematic noise patterns \citep[see][]{thalmann2013} than the locally optimized combination of images \citep[LOCI,][]{lafreniere2007} and seems to yield similar S/N images \citep[e.g.][]{meshkat2013}. We first cropped the images and kept the central 600 $\times$ 600\,px = 5.7$\arcsec$ across,  while blocking the stellar saturated pixels. Then we subtracted the mean of the image stack from each frame. For PCA the data is used to obtain the orthogonal basis of eigenvectors that form the different eigeinmages or modes. The projected image onto each orthogonal mode is finally subtracted from the frame. The ADI approach is completed by derotating all the images according to their individual parallactic angle and collapsing them together into one single final PCA-ADI image for every subtracted mode. The mean is used in our reduction instead of the median to assure linearity of the data  \citep{thalmann2013,brandt2013}. After visual inspection of the output images, we concluded that the subtraction of five modes out of 53 provides a good balance between flux self-subtraction and speckle noise suppression. Moreover, we also undertook the classical ADI reduction complemented with LOCI (rotation gaps N$\delta$ = 0.75 and optimization areas NA = 300 in units of the PSF FWHM),  where the parameters are optimized to maximize the sensitivity to the faint disk. 

\subsection{Polarized differential imaging }
The PDI dataset was reduced following the usual SEEDS polarimetry procedure \citep[e.g.][]{hashimoto2011}. Each exposure was first destriped, flat-field and dark corrected and then cleaned for hot and bad pixels. Next, we removed the distortions generated by the inclusion of the Wollaston prism. Both polarization directions in each exposure were cut out and all images aligned before the Stokes parameters $Q$ and $U$ were retrieved after the subtraction of each pair of perpendicular flux polarimetric components \citep[see double-difference technique;][]{hinkley2009}. We finally accounted for instrumental polarization \citep{joos2008}.

The conventional way of calculating the linear polarized intensity from the squared Stokes parameters $PI=\sqrt{Q^{2} + U^{2}}$ carries a halo of large positive systematic errors as a result of the $Q$ and $U$ addition in quadrature.  For an optically thin disk under the single-scattering assumption, only linear polarization in the azimuthal direction is expected. In this way, when considering polarized intensity rather than total intensity, a single-scattering approximation is appropriate, since multiple scattering randomizes the scattering angle, resulting in a depolarization of multiply scattered photons. Therefore, it is beneficial to transform the regular Cartesian-coordinate Stokes Parameters into a polar coordinate system ($Q_{\phi}, U_{\phi}$), where the azimuthally polarized light (perpendicular polarization with respect to the line between the star and a given point in the image) appears as a positive signal in the $Q_{\phi}$ image \citep[see][]{benisty2015, schmid2006}. On the contrary, $U_{\phi}$ is free of signal and can be used as an estimation of the noise present in the $Q_{\phi}$ image \citep[][]{avenhaus2014}.

\section{Results and discussion}

\subsection{ADI-processed disk images}
The debris disk in scattered light around HD 32297 is clearly seen in the final ADI-treated images and S/N maps: Figure \ref{disk}. The maps are first convolved with a 5-pixel diameter ($\sim$FWHM) circular aperture to determine the S/N per resolution element, followed by the creation of a noise map from the standard deviation of pixels forming concentric annuli from the star. The region where the disk is present was avoided when constructing these noise maps.  We are able to detect the disk at significance levels of $\ge$\,5$\sigma$ from $\sim$0.45$\arcsec$ out to $\sim$1.7$\arcsec$ (50--192\,AU) in projected separation from the star. 

Although the saturation radius was only 12 px = 0.11$\arcsec$, the speckle noise dominates the image at small separations, and thus no information can be recovered up to $\sim$50\,AU from the star, probably due to a poor rotation range for these data. A higher field rotation would in principle have implied  a higher ADI quality and thus an inner radius in which the disk dominates. However, previous $H$-band ADI works \citep{boccaletti2012,esposito2014} show a comparable effective inner working angle\footnote{The IWA typically refers to the separation limit set by a coronagraph. However, for these observations we did not use any coronagraph, but rather we define an effective IWA corresponding to the smallest separation from the star at which the disk can be detected.} (IWA) with a higher change in parallactic angle (see Table \ref{table1}). This is probably because of the more conservative approach used by both of these groups, which reduces self-subtraction, but the disadvantage comes in the form of a poorer noise attenuation and a bigger IWA.  As the PCA-ADI technique demonstrates a better disk flux conservation for these data, we continue to perform our scientific analysis on the PCA image only. Certainly, a rather high Strehl ratio in our SEEDS data ($\sim$0.4) and good seeing results in less energy in the broad PSF halo and excellent PSF stability. This, together with the PCA-ADI reduction, culminate in a competitive flux recovery and a favourable S/N image when comparing our observations with previous works.

The disk is nearly edge-on and shows a largely symmetrical structure in brightness and size with negative-brightness regions surrounding the disk from the PCA-ADI algorithm causing self-subtraction.  However, we note two striking morphological features. First, the disk exhibits a slight deviation from the midplane in agreement with all previous ADI-based studies in $H$, $K$, and $L$ bands \citep[e.g.][]{rodigas2014,esposito2014,currie2012}. This is most likely because the disk is not perfectly edge-on, forming an elongated ellipse shape, in combination with forwards scattering grains, which would make the region of the disk between us and the star appear brighter than the opposite side. \citet{boccaletti2012} first dedicated their $H$- and $K$-band study to explain this morphology by means of GRaTer \citep{augereau1999}  scattered-light disk models and found that the performed minimizations pointed to a considerably anisotropic scattering factor and a clear inclination of 88$^{\circ}$ in both bands. The same inclination was also found by Currie et al. (2012) in the $K$ band, who also produced several dust models and showed that the global morphology is recreated well by strong forwards scattering grains close to the star and weakly forwards scattering grains at large projected separations. The SB profiles, however, were also reproduced with other parameters and combinations of models. 

Moreover, we find that the lobes present an approximate, symmetrical flux depression, which are indicated with two arrows in Figure \ref{disk}(a) at distances of $\sim$0.65$\arcsec$ (SW lobe) and $\sim$0.75$\arcsec$ (NE lobe). These depressions, or potential gaps, are more clearly discernible in the LOCI reduction, as in our case is more aggressive than PCA in terms of flux conservation. However, the SB profile computed from the PCA-reduced data (see Section 4.1.2) also shows dips at exactly the same distances as the LOCI images, which favours this interpretation. Yet, this result has to be taken with caution, as this is the first time such a structure has been detected in the HD 32297 disk, while other works with better field rotation (as \citet{esposito2014} and \citet{boccaletti2012}, see Table \ref{table1})  and higher Strehl ratio 
have not been able to observe a depression, but only turnovers or a wavy structure either in the images or in the inferred SB profiles. The discernibility of the mentioned structure by our side might be due to the combination of a good seeing and a competitive Strehl ratio performance, which prevents contamination of light from the seeing halo of surrounding regions of the disk into the cavity. In the following, we model the disk morphology as seen from the reduced PCA-ADI image and study its polarized structure.

\begin{figure*}
\centering
\includegraphics*[scale=0.35]{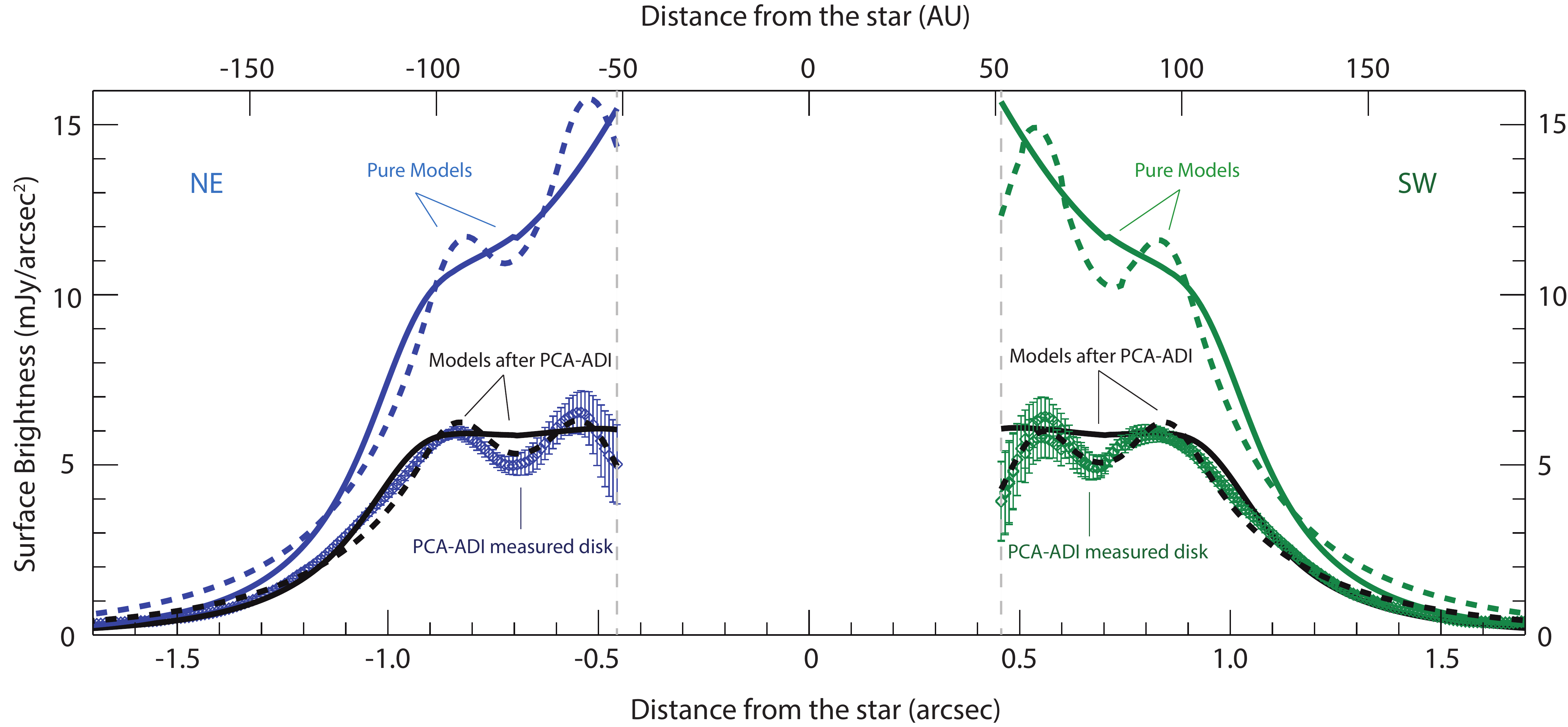}
\caption{HD 32297 full intensity SB profiles measured in circular apertures of 6 px ($\simeq$ 0.06$\arcsec$) centred on the brightest pixel at each radial separation from the star for the NE and SW sides. The dotted profiles with error bars are obtained from the ADI-PCA reduced data. The errors represent the standard deviation of SB measurements covering azimuthal angles where no disk flux is present. Thick curves indicate the best-fit single model, before (pure models) and after (models after PCA-ADI) undergoing the ADI-PCA reduction. Dashed curves show the same approach but for the best-fit NE and SW double models (see Table \ref{table3}).}
\label{sb}
\end{figure*}

\subsubsection{Position angle}
To begin with, we determine the position angle (PA) that best accounts for the global structure of the disk, taking into account that the spine of the disk changes with radial separation from the star, as seen in the data. We followed the procedure presented in \citet{thalmann2013} for HIP 79977. The image is first de-rotated by an angle that we estimate will leave the disk horizontal in the frame. Then the disk is mirrored about  the y-axis and subtracted from the unmirrored image. This residual image is then binned by a factor of 5 ($\sim$FWHM), so each resulting pixel represents its own independent resolution element. A noise map profile is taken as the standard deviation of the pixel values in concentric annuli from the HD 32297 star, but avoids an evaluation region formed by the pixels in which disk flux is present. Finally, we obtain a S/N map by dividing the residual image by the noise profile, from which the $\chi^{2}$ value is calculated as the sum of the squared values of all the pixels that the evaluation region encompasses. This process is repeated for position angles of [47.0, 47.1...48.4, 48.5] degrees and results in a minimum $\chi^{2}_{min}$ at PA = 47.8 $\pm$ 0.1$^{\circ}$, eastwards of north. This is consistent with previous values reported by HST and ADI works imaging close regions from the star, both from actual data measurement and modelling results (e.g. \citet{debes2009}, 46.3 $\pm$ 1.3$^{\circ}$; \citet{boccaletti2012}, 47.4 $\pm$ 0.3$^{\circ}$; \citet{esposito2014}, 47.5$^{\circ}$).

\subsubsection{Surface brightness profile}

We measure the SB profile of the debris disk directly from the reduced images after they are processed by the PCA-ADI technique. We first rotate the disk by 90$^{\circ}$-PA counterclockwise, thereby leaving the midplane approximately horizontal. The stellar PSF in the non-saturated frames is used to flux calibrate the reduced saturated image. To this end, every pixel in the disk image is interchanged by the total number of counts in an aperture of 5 px ($\sim$FWHM) centred on itself, forming a single resolution element that is then converted to mJy/arcsec$^{2}$. Next, the disk is isolated in a narrow slit and we determine the brightest pixel at each distinct distance in integer units from the star. A circular aperture with a radius of 6 px $\simeq$ 0.06$\arcsec$ is centred on each of the brightest elements and the embraced pixel values are added together. This aperture is chosen so as to enclose all the pixels containing disk flux, and at the same time, to avoid including background contamination.  When pixels with negative values are incorporated inside the aperture, they are treated as not being part of the disk and their values are changed to zero. At a given radius, the uncertainty is calculated from the standard deviation of SB measurements inside an aperture equal to that used for the disk SB, but covering azimuthal angles where disk flux is not present. 

The results are shown in Figure \ref{sb}(dotted profiles with error bars). The disk brightens all the way down to $\sim$0.85$\arcsec$. From here to closer separations, we clearly see the presence of two different so-called bumps in both lobes of the disk: the outer one peaks at $\sim$0.80$\arcsec$, while the inner increment peaks at $\sim$0.55$\arcsec$, both in projected distance.  The intermediate valley reproduces the likely gap profile seen in the actual image at distances of  $\sim$0.65$\arcsec$ and $\sim$0.75$\arcsec$ for the SW and NE lobe, respectively, which is the distance at which the two arrows in Fig \ref{disk} indicate the flux depression. The fact that the same global structure is observed on both sides is a strong argument in favour of the gapped scenario, which is also supported by the bumps consistency within the determined error bars. As different teams use different procedures and apertures, we know that the selection of aperture size has a clear impact on the SB \citep[see e.g.][]{esposito2014}. As a matter of prudence, we verified that these dips are always detected with all sensible aperture size choices. 

The SW-NE sides are essentially symmetric, but the NE lobe appears to be slightly displaced ($\sim$FWHM) outwards compared to the SW ansa. \citet{boccaletti2012} also found a similar shift from a more flux conserving classical ADI reduction data. Towards bigger radii, we find breaks in the power laws at $\sim$95 and $\sim$130\,AU, which agrees perfectly with the \citet{esposito2014} $H$-band SB truncation distances.  On the other hand, our work is the first that displays such a gapped interior structure from $\sim$0.9$\arcsec$ down to $\sim$0.6$\arcsec$.  From the interior to the two peaks, the brightness decreases again inwards from $\sim$0.55$\arcsec$ in agreement with previous $H$-band results, although the speckle noise begins to dominate and hinders the analysis. However, a different trend is seen in the $K$ and $L$ bands \citep[e.g.][]{rodigas2014} where the profiles continue rising sharply.

\begin{figure*}
\centering
\includegraphics*[scale=0.5]{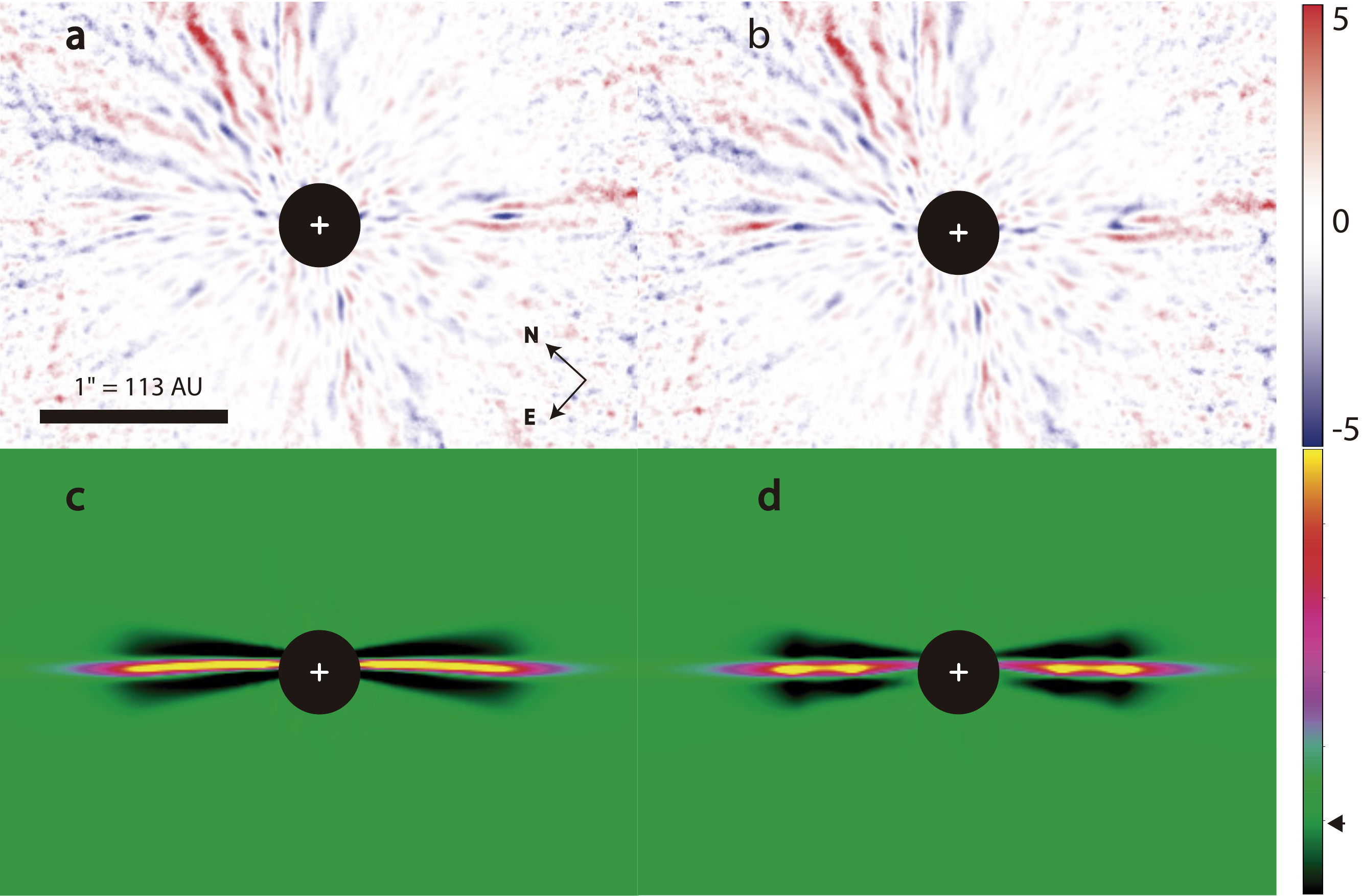}
\caption{ HD 32297 debris disk global minimization modeling. \textbf{(a)} Residual map in a  [-5$\sigma$, 5$\sigma$] linear stretch after subtracting the best-fit single model after PCA-ADI reduction. \textbf{(b)} The same as \textbf{(a)} after subtracting the best-fit double model (see Table \ref{table2}). \textbf{(c)} Best-fit single model after undergoing PCA-ADI reduction, in linear scale and not convolved. The arrow in the colour bar indicates the level zero in counts: black shows over-subtracted regions. \textbf{(d)} The same for the best-fit double model. The images have not been binned and the diameter of the black mask is $\sim$0.4$\arcsec$.} 
\label{residuals}
\end{figure*}

\subsubsection{Single ring models}
The major detriment to ADI is the flux and morphological loss on disks \citep{milli2012}. To understand the original disk structure and its grain distribution, we explore a grid of scattered light models and try to forwards model the effects that the PCA-ADI reduction had in the real image.  By means of the GRaTer code \citep{augereau1999}, we generate ring models that undergo the PCA-ADI procedure in the same way as the actual disk. That is, we project each parallactic-rotated model frame onto the first five modes or eigen images delivered by the real data, and these projected images are then subtracted from their corresponding model frame.

The model parameters that are varied and define the disks are the ring semi-major axis \textit{r$_\mathrm{0}$}, the power-law slopes of the radial brightness distribution in the inner and outer region of the disk $\alpha_\mathrm{in}$ and $\alpha_\mathrm{out}$, the inclination \textit{i} and the Henyey-Greenstein scattering function \textit{g}. When minimizing the residual $\chi^{2}$  as defined in section 4.1.1, we determine the global scale factor between the model and the data as well as the image shift in the y-direction that gives the best fit. The latter accounts for an inaccurate absolute centring of the stellar PSF in the reduction process. The position angle was fixed to 47.8$^{\circ}$, as previously determined, and the eccentricity was set to zero. We note that the models are symmetrical, so our procedure assumes that both sides of the disk image are identical, which is not strictly the case as seen in Figure \ref{disk}.

A total of 240 models with scattered light are created, which span a range of parameters based on previous modelling work carried out on the HD32297 disk \citep{boccaletti2012,currie2012} with \textit{r$_\mathrm{0}$} = [100, 110, 120]\,AU,  $\alpha_\mathrm{in}$ = [2, 5, 8, 10], $\alpha_\mathrm{out}$ = [-2, -4, -6, -8], and \textit{g} = [0.1, 0.3...0.9]. The inclination was set to \textit{i} = 88$^{\circ}$. Our global minimization delivers a best-fit model with all parameters enclosed by the selected range (i.e. no boundary values), \textit{r$_\mathrm{0}$} = 110\,AU, $\alpha_\mathrm{in}$ = 8,  $\alpha_\mathrm{out}$ = -6, \textit{g} = 0.5 and a y-offset of 3 pixels NW for the disk image. The attained minimum is $\chi^{2}_\mathrm{min}$ = 958 = 1.9*Ndata, where $Ndata$ is the number of binned pixels enclosed by the evaluation region. This model matches perfectly the global best-fit solution found by \citet{boccaletti2012} also with the GRaTer code and in the $H$ band; all of the parameters coincide exactly except for a somewhat steeper inner slope in their case. As our reduction does not provide fully independent resolution elements even after binning, we show a range of well-fitting models matching the data within a threshold of $\sqrt{2N_\mathrm{data}}$ \citep[same as][]{thalmann2013} from the best-fit solution. We have ten models that fall within the parameters shown in Table \ref{table2}. All of these present a high forwards scattering function of \textit{g} = 0.5, while the inner slope is not well constrained. We show the residual S/N map of the disk after subtraction of the best-fit model in Figure \ref{residuals}(a). The model appears successful in representing the global morphological features of the data down to the speckle noise level and also reproduces the characteristic small-scale bow shape (see panel (c)). 

This model also provides the best-fit solution if the criterion for a model selection relies on a SB minimization after PCA-ADI correction in the range 0.45$\arcsec$--1.7$\arcsec$. In this approach we treat and fit the selected range of models to the NE and SW lobes separately. The SB of the best single model after PCA-ADI reduction is thus plotted in Figure \ref{sb} as a thick black curve. The corresponding pure model is shown as a thick blue curve. Given that the models succeed in recreating the observed disk structure, pure model profiles represent the true SB of the disk, as they are computed before going through an ADI reduction and thus are not affected by flux and morphological loss. An unambiguous outcome from these profiles is that the simple single ring model cannot account for the bumpy profile of the real disk. Its strong forwards scattering Henyey-Greenstein function (\textit{g} = 0.5) allows us to reproduce the drop off at large distances perfectly, but the model is incapable of recreating the gapped structure interior to $\sim$0.80$\arcsec$, adopting a medium flat value that works well when minimizing the residuals. We checked that no single scattering factor replicates the sharp inner gap; models smoothly scatter more or less light than the actual disk at small distances depending whether g$>$0.5 or g$<$0.5, respectively. 

As we are conducting a qualitative analysis, we avoid the choice of a restrictive cut-off when determining the range of well-fitting parameters, i.e, those that yield an acceptable fit to the SB profile. Instead, we define the set of good-fitting models as those that meet the condition $Res < ( Res_\mathrm{min} + \sqrt{2 \Delta_\mathrm{Res}} ) $, where $\Delta_\mathrm{Res} $ is the difference in the SB residuals of the SW and NE best-fit models, and $Res_{min}$ is taken as the residual of the best-fit NE model. As in the case of the global minimization, medium to strong forwards scattering grains are favoured, and the inner slope does not appear to have an impact on the final solution (see Table \ref{table3}).

 \begin{table}
\caption{Best model parameters from the global $\chi^{2}$ minimization}
\label{table2} 
\centering
\begin{tabular}{c c c }
\hline\hline
Parameter & \textbf{Best-fit single model} & Well fitting range \\ 
\hline 
$r_\mathrm{0}$ (AU) &110&100--120\\
$\alpha_\mathrm{in}$&8.0&2.0--10.0\\
$\alpha_\mathrm{out}$ &-6.0&-4 -- -6\\
g &0.5&0.5\\
 i  ($^{\circ}$)&88.0&fixed\\
\hline
& \textbf{Best-fit double model} &  \\ 
\hline 
$r_\mathrm{0in}$ (AU) &60&fixed\\
$r_\mathrm{0out}$ (AU) &90&fixed\\
$\alpha_\mathrm{in}$&40.0&15.0--40.0\\
$\alpha_\mathrm{out}$ &-5.0&-4 -- -5\\
g &0.2&0.2--0.3\\
 i  ($^{\circ}$)&88.0&fixed\\
Scale factor &2.0&1.0--2.0\\
\end{tabular}
\end{table}

 \subsubsection{Double ring models}
In an attempt to reproduce the gapped structure, we refine the analysis by creating two-ring models, where the rings are located at the positions of the observed peaks in the SB profile. Other than the radius, we consider all the parameters of both rings to be equal in each model. This is motivated by several factors. First, if the two rings are the source of the observed scattered light, the grains probably formed beyond the H$_\mathrm{2}$0 and CO$_\mathrm{2}$ snow lines and around the CO line location \citep[][]{qi2015}, so most of these compounds should have condensed into solid grains and the composition of both rings should be similar. This would in principle imply a similar Henyey-Greenstein scattering function value. Secondly, the possible presence of a second unresolved inner dust component has been a frequent claim made by several groups, especially from SED modelling. \citet{fitzgerald2007} found that the fit is improved if a population near $\sim$60\,AU accounts for the 60\,$\mu$m emission. More than one population was also a clear outcome from the \citet{currie2012} SED modelling (0.4-1300\,$\mu$m), suggesting an inner component at $\sim$45\,AU. Even the addition of a third dust population very close to the star delivered a very good fit to the data if the rings are located at the locations of the SB peaks (45 and 110\,AU for their $K$-band data). With new far-IR data, Donaldson et al. (2013) also required an extra warm component, as an inner disk near the stellar habitable zone. Furthermore, SB rising profiles at close separations in the $J$, $K$, and $L$ bands \citep[e.g.][]{rodigas2014} allows us to infer the existence of a ring interior to the IWA. Finally, considering the same parameters for the two disks reduces the computational effort noticeably.

We place the inner and outer ring at 60\,AU and 95\,AU, respectively. As we see in the SB profile, the sharp peaks at those distances require weakly or moderately forwards scattering grains and steep slopes, otherwise the scattering would be much higher close to the star and the profiles would not peak at the desired projected distances, so we create models with  $\alpha_\mathrm{in}$ = [5, 10, 15, 40] (where 40 simply represents a sharp cut-off) , $\alpha_\mathrm{out}$ = [-3, -4, ..., -7] and g = [0.0, 0.1, ..., 0.6]. The inclination and position angle are again fixed as before. Here we also include as a parameter a density scale factor that makes the outer disk more populated and accounts for the contrast in brightness between the rings, creating 701 double ring models in total.

A global minimization delivers a best-fit model with $\chi^{2}_\mathrm{min}$ = 1003 = 2*Ndata, which is shown in Figure \ref{residuals}(d) together with its residual image (Figure \ref{residuals}(b)). This double model achieves a worse $\chi^{2}$ minimization compared to the best single-ringed model and falls outside its range of well-fitting models. This might be due to the low $g$ value obtained (see Table \ref{table2}), as the bow shape may be more difficult to emulate (see panels (c) and (d) in Figure \ref{residuals}). We forced both rings in the double models to be equal; greater freedom in this regard would likely achieve a better global fit while still reproducing the gapped profile.  In any case, double models result in a manifestly better overall SB recreation than the single ring models (especially its bumpy structure), shown as dashed curves in Figure \ref{sb}. Thus, only very sharp double ring disks with weakly forwards scattering grains can account for the gap feature at such stellar distance, which would consist of the less dust abundant region between the disks. In this way, from both approaches we get a well-fitting range of double models consisting of weakly forwards scattering (g $\leq$\,0.3) grains (see Table \ref{table2} and Table \ref{table3}).  Again the inner slopes are basically unconstrained, while all but one of the models favour a more dense outer ring, 55$\%$ and 52$\%$ of them have between 1.5--1.75 times more dust than their inner counterpart, respectively, for the SB and global minimizations.

 \begin{table}
\caption{Best model parameters from the surface brightness minimization}
\label{table3} 
\centering
\begin{tabular}{c c c}
\hline\hline

\tiny{Parameter} & \tiny{\textbf{NE / SW Best-fit single model}}  & \tiny{Well fitting range} \\ 
\hline 
\tiny{$r_\mathrm{0}$ (AU)} &\tiny{110} / \tiny{110}&\tiny{100--120}\\
\tiny{$\alpha_\mathrm{in}$} &\tiny{8.0} / \tiny{8.0}&\tiny{2.0--40.0}\\
\tiny{$\alpha_\mathrm{out}$} &\tiny{-6} / \tiny{-6}&\tiny{-4 -- -8}\\
\tiny{g} &\tiny{0.5} / \tiny{0.5}&\tiny{0.3--0.5}\\
 \tiny{i  ($^{\circ}$)} &\tiny{88.0} / \tiny{88.0}&\tiny{fixed}\\
\hline

\tiny{} & \tiny{\textbf{NE / SW Best-fit double model}}  & \tiny{} \\ 
\hline 
\tiny{$r_\mathrm{0in}$ (AU)} &\tiny{60} / \tiny{60}&\tiny{fixed}\\
\tiny{$r_\mathrm{0out}$ (AU)} &\tiny{95} / \tiny{95}&\tiny{fixed}\\
\tiny{$\alpha_\mathrm{in}$} &\tiny{40.0} / \tiny{40.0}&\tiny{15.0--40.0}\\
\tiny{$\alpha_\mathrm{out}$} &\tiny{-4} / \tiny{-4}&\tiny{-4}\\
\tiny{g} &\tiny{0.2} / \tiny{0.1}&\tiny{0.0--0.3}\\
 \tiny{i  ($^{\circ}$)} &\tiny{88.0} / \tiny{88.0}&\tiny{fixed}\\
\tiny{Scale factor} &\tiny{1.50} / \tiny{1.50}&\tiny{1.25--2.0}\\
\hline
\end{tabular}
\end{table}

\subsection{Polarimetric Disk images}
  
\begin{figure*}
\centering
\includegraphics*[scale=0.5]{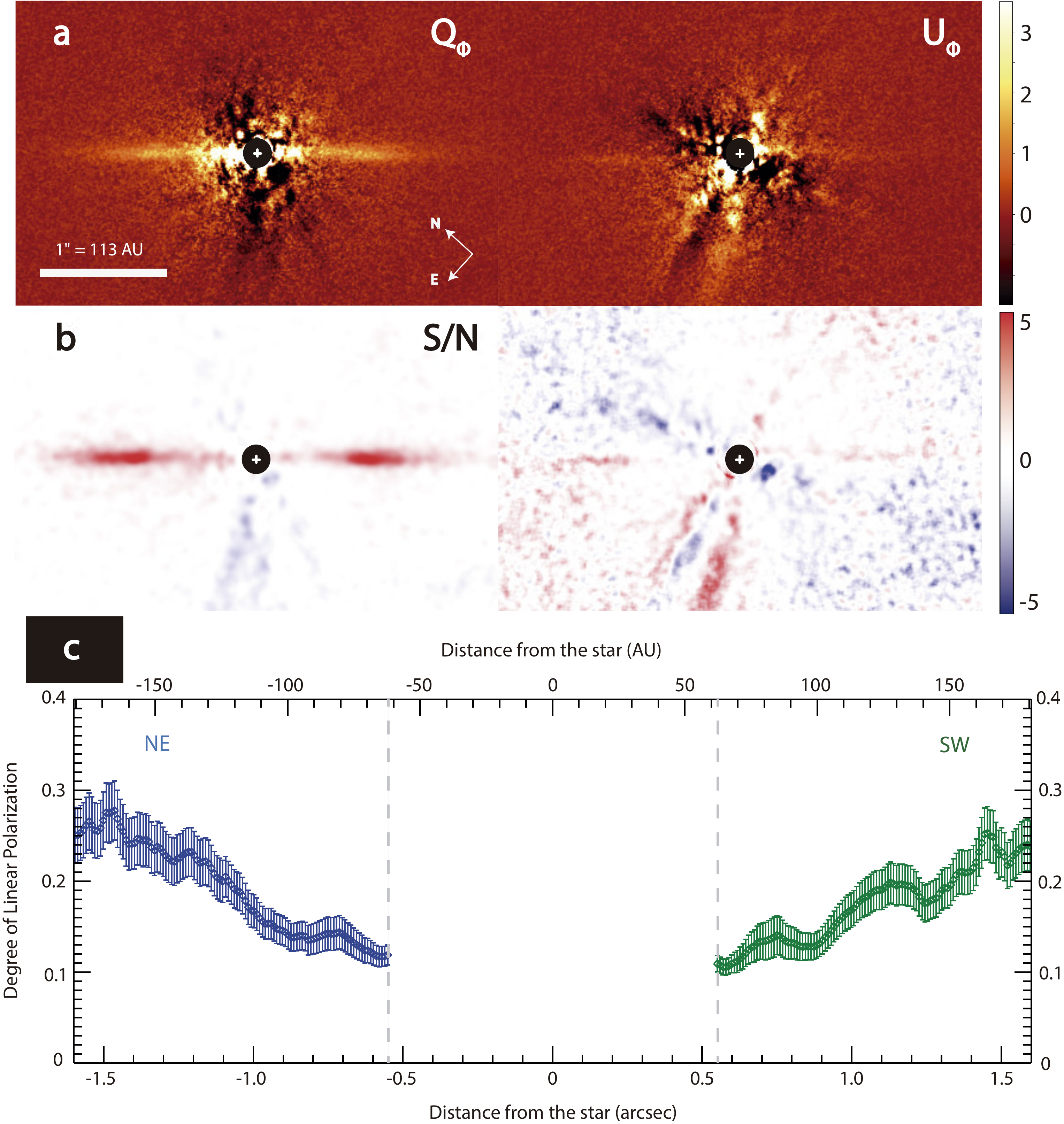}
\caption{ HD 32297 Subaru/HICIAO polarized differential imaging (PDI) results in the $H$ band. \textbf{(a)} $Q_{\phi}$ and $U_{\phi}$ Stokes parameters obtained from the 2014 October 10 data in linear stretch and units of mJy/arcsec$^{2}$. We recover the deviation from the midplane seen in the ADI data (see Figure \ref{disk}), but do not observe any gapped structure. The $U_{\phi}$ image has no clear structure apart from a very faint signal in the midplane of the disk, which is likely due to an imperfect polarimetric calibration. \textbf{(b)} Corresponding S/N images in the linear [-5$\sigma$, 5$\sigma$] scale. The images have been convolved with a circular aperture of diameter $\sim$FWHM. The signal in the $Q_{\phi}$ S/N image (left panel)  is not significant at small distances from the star. The noise map was obtained as the standard deviation of the $U_{\phi}$ image (right panel) in star-centred annuli.  \textbf{(c)} Linear polarization degree as a function of the projected distance from the star for both sides of the disk. The final polarization degree increases from $\sim$10$\%$ at $\sim$0.55$\arcsec$ to $\sim$25$\%$ at $\sim$1.6$\arcsec$, as at larger separations the scattering angles are constrained to $\sim$90$^{\circ}$, where the highest amount of polarization is produced.} 
\label{polarimetry}
\end{figure*}

The HD 32297 scattered light $Q_{\phi}$ and $U_{\phi}$ Stokes parameters are shown in Figure \ref{polarimetry}(a). Under the assumption of optically thin disk and single scattering, only linear polarization in the azimuthal direction is expected, which should show up in $Q_{\phi}$  as a positive signal; $U_{\phi}$ retains the polarization in the radial direction and in principle contains only noise. The $Q_{\phi}$ image retrieves the deviation from the midplane already seen in the ADI-treated data, although in this case no gapped structure can be discerned. The residual $U_{\phi}$ image predominantly presents noise and a very faint structure following the midplane of the disk. This could be an indication of a change in the polarization angle, but based on the very weak signal, we consider an imperfect calibration to be the most likely cause for this effect \citep[see also][]{avenhaus2014}.

We show in Figure \ref{polarimetry}(b) the S/N maps of the $Q_{\phi}$ and $U_{\phi}$ images in a [-5$\sigma$, 5$\sigma$] linear stretch, where we used the standard deviation of the $U_{\phi}$ parameter in concentric annuli as an estimation of the noise for the $Q_{\phi}$ and $U_{\phi}$ pixel values in the same annuli \citep[][]{thalmann2015}. The disk is detected extending to $\sim$1.6$\arcsec$ ($\sim$170\,AU) and down to $\sim$0.25$\arcsec$ ($\sim$28\,AU), while achieving S/N values above 5$\sigma$ over the range 0.75--1.2$\arcsec$. As seen in the  $Q_{\phi}$ S/N map (Figure \ref{polarimetry}(b), left panel), the signal is not significant at small projected distances. We thus decided not to consider pixel values inside a radius of $\sim$0.55$\arcsec$ from the star for further analysis.

The radial surface polarized brightness profile of the disk is derived from the $Q_{\phi}$ image. We isolate the disk and use the positions of the brightest pixels of the best-fit double ring model as a centre of an aperture of 6\,px $\simeq$\,0.06$\arcsec$ in radius. Afterwards, this radial profile is divided by the full intensity radial profile of the SB-minimization best double ring model for each side (i.e. before PCA-ADI), calculated in the same positions, to finally obtain the linear polarization degree $P$ (Figure \ref{polarimetry}(c)). Error bars are obtained by adding in quadrature $U_{\phi}$ errors and the standard deviation of polarization degree measurements of the well-fitting double models for each side of the disk (see Table \ref{table3}). The adoption of the double ring model is motivated by its ability to recreate the gapped structure of the observed disk. 

The polarization degree increases in a rather symmetric way on both sides of the disk from $\sim$10$\%$ at $\sim$0.55$\arcsec$ to $\sim$25$\%$ at $\sim$1.6$\arcsec$. This is the same behaviour as found by \citet{thalmann2013} for HIP 79977, but in that case the dust grains favoured a higher polarization degree, reaching $\sim$45$\%$ at 1.5$\arcsec$. \citet{min2012} modelled polarimetric images of protoplanetary disks and obtained the degree of linear polarization of the dust particles as a function of the scattering angle (see Figure 1 in their paper). The highest amount of polarization comes from position angles in the range 70--90$^{\circ}$. \citet{perrin2015} also computed polarization curves for a single silicate composition to explain their HR 4796A observations (see Figure 14 in their paper), finding similar results. This might explain why our polarization degree increases at larger separations from the star, as the scattering angles are increasingly constrained to $\sim$90$^{\circ}$, whereas close to the star small scattering angles reduce the polarization fraction.

Given the almost edge-on geometry of the HD 32297 debris disk, we cannot compute clean polarization degree values, as different locations in the disk contribute to the observed polarization signal at every distance from the star. We can, however, assume that the edge of our modelled outer ring located at 95\,AU scatters off stellar light at $\sim$90$^{\circ}$ and no other dust particles are affecting the measurement. From Figure \ref{polarimetry}(c) we obtain a polarization degree of $\sim$15$\%$ at the location of the outer ring, and a polarization degree that is higher outwards of the ring. This might be because smaller dust particles are blown out from the ring, as very small grains tend to be more polarized.  In any case, a polarization degree of $\sim$15$\%$ at a scattering angle of $\sim$88$^{\circ}$ can be qualitatively interpreted from \citet{perrin2015} models as grains of radius 0.4--0.8\,$\mu$m. \citet{min2012} models favour a structure of compact homogeneous particles rather than fluffy inhomogeneous aggregates. These grains would be blown out to larger radii from the outer ring by the stellar wind and would scatter off starlight at $\sim$90$^{\circ}$, which could explain why the polarization degree profile keeps increasing outside of the location of the ring radius.

\section{Conclusions}

The unpolarized observations show a gapped structure, which cannot be regarded as categorical as other ADI studies with better field rotation \citep[e.g.][]{esposito2014} or higher Strehl ratio \citep[e.g.][]{currie2012} were not able to detect a gap in their images. Our outcome might be due to better seeing conditions during the observations, which avoided contamination of light from the seeing halo of surroundings regions of the disk into the gap. This finding would, in principle, agree with several SED modelling studies that have claimed the presence of two dust components at different locations \citep[e.g.][]{donaldson2013,currie2012,fitzgerald2007}, or even with rising SB profiles at close stellar separations \citep[e.g.][]{rodigas2014}. The gap would then represent the intermediate region between both rings. One should keep in mind that our models are very simple and we do not aim to recreate the disk structure perfectly, but to demonstrate that a double ring (or a more complicated structure) could actually be a plausible explanation for the observations. Double-ringed disks are emerging as relatively common, and have been observed in several systems, such as HD 141569 \citep[e.g.][]{biller2015}, HD 107146, and HD 92945, where double-ringed disks could be produced by unseen eccentric planets \citep[][]{pearce2015}. In this way, a simple ring model cannot be discarded either, because it produces better global-fit residuals, albeit being not able to replicate the SB profile.

The first polarimetry observation of the HD 32297 disk does not present any evidence for a gapped scenario, although there could be several explanations for this absence, such as low S/N values at the gaps separation or a change in the degree of linear polarization of the dust grains as a function of separation from the star. We find that the polarization of the grains increases all the way  from $\sim$10$\%$ at 0.55$\arcsec$ to $\sim$25$\%$ at 1.6$\arcsec$. The maximum is then found where scattering angles are $\sim$90$^{\circ}$, which agrees well with linear polarization models as a function of scattering angle \citep[e.g.][]{min2012}, where the highest polarization comes from angles between 70--90$^{\circ}$. The dust size, structure, and composition is rather complicated to unveil given the edge-on geometry of the disk that prevents us from computing a clear polarization degree for a range of scattering angles. From the \citet{perrin2015} and \citet{min2012} models, we were able to speculate on 0.5\,$\mu$m compact, homogenous grains that are blown out to larger radii by the solar wind.

\begin{acknowledgements}
We would like to thank J.C. Augereau for providing the GraTeR code used to create our disk models. R. Asensio-Torres and M. Janson gratefully acknowledge funding from the Knut and Alice Wallenberg foundation. J. Carson acknowledges support via the U.S. National Science Foundation under Award No. 1009203.
\end{acknowledgements}

\bibliography{ref}

\begin{thebibliography}{47}
\expandafter\ifx\csname natexlab\endcsname\relax\def\natexlab#1{#1}\fi

\bibitem[{{Akiyama} {et~al.}(2015){Akiyama}, {Muto}, {Kusakabe}, {Kataoka},
  {Hashimoto}, {Tsukagoshi}, {Kwon}, {Kudo}, {Kandori}, {Grady}, {Takami},
  {Janson}, {Kuzuhara}, {Henning}, {Sitko}, {Carson}, {Mayama}, {Currie},
  {Thalmann}, {Wisniewski}, {Momose}, {Ohashi}, {Abe}, {Brandner}, {Brandt},
  {Egner}, {Feldt}, {Goto}, {Guyon}, {Hayano}, {Hayashi}, {Hayashi}, {Hodapp},
  {Ishi}, {Iye}, {Knapp}, {Matsuo}, {Mcelwain}, {Miyama}, {Morino},
  {Moro-Martin}, {Nishimura}, {Pyo}, {Serabyn}, {Suenaga}, {Suto}, {Suzuki},
  {Takahashi}, {Takato}, {Terada}, {Tomono}, {Turner}, {Watanabe}, {Yamada},
  {Takami}, {Usuda}, \& {Tamura}}]{akiyama2015}
{Akiyama}, E., {Muto}, T., {Kusakabe}, N., {et~al.} 2015, \apjl, 802, L17

\bibitem[{{Augereau} {et~al.}(1999){Augereau}, {Lagrange}, {Mouillet},
  {Papaloizou}, \& {Grorod}}]{augereau1999}
{Augereau}, J.~C., {Lagrange}, A.~M., {Mouillet}, D., {Papaloizou}, J.~C.~B.,
  \& {Grorod}, P.~A. 1999, \aap, 348, 557

\bibitem[{{Avenhaus} {et~al.}(2014){Avenhaus}, {Quanz}, {Schmid}, {Meyer},
  {Garufi}, {Wolf}, \& {Dominik}}]{avenhaus2014}
{Avenhaus}, H., {Quanz}, S.~P., {Schmid}, H.~M., {et~al.} 2014, \apj, 781, 87

\bibitem[{{Benisty} {et~al.}(2015){Benisty}, {Juhasz}, {Boccaletti},
  {Avenhaus}, {Milli}, {Thalmann}, {Dominik}, {Pinilla}, {Buenzli}, {Pohl},
  {Beuzit}, {Birnstiel}, {de Boer}, {Bonnefoy}, {Chauvin}, {Christiaens},
  {Garufi}, {Grady}, {Henning}, {Huelamo}, {Isella}, {Langlois}, {M{\'e}nard},
  {Mouillet}, {Olofsson}, {Pantin}, {Pinte}, \& {Pueyo}}]{benisty2015}
{Benisty}, M., {Juhasz}, A., {Boccaletti}, A., {et~al.} 2015, \aap, 578, L6

\bibitem[{{Biller} {et~al.}(2015){Biller}, {Liu}, {Rice}, {Wahhaj}, {Nielsen},
  {Hayward}, {Kuchner}, {Close}, {Chun}, {Ftaclas}, \& {Toomey}}]{biller2015}
{Biller}, B.~A., {Liu}, M.~C., {Rice}, K., {et~al.} 2015, \mnras, 450, 4446

\bibitem[{{Boccaletti} {et~al.}(2012){Boccaletti}, {Augereau}, {Lagrange},
  {Milli}, {Baudoz}, {Mawet}, {Mouillet}, {Lebreton}, \&
  {Maire}}]{boccaletti2012}
{Boccaletti}, A., {Augereau}, J.-C., {Lagrange}, A.-M., {et~al.} 2012, \aap,
  544, A85

\bibitem[{{Boccaletti} {et~al.}(2015){Boccaletti}, {Thalmann}, {Lagrange},
  {Janson}, {Augereau}, {Schneider}, {Milli}, {Grady}, {Debes}, {Langlois},
  {Mouillet}, {Henning}, {Dominik}, {Maire}, {Beuzit}, {Carson}, {Dohlen},
  {Engler}, {Feldt}, {Fusco}, {Ginski}, {Girard}, {Hines}, {Kasper}, {Mawet},
  {M{\'e}nard}, {Meyer}, {Moutou}, {Olofsson}, {Rodigas}, {Sauvage},
  {Schlieder}, {Schmid}, {Turatto}, {Udry}, {Vakili}, {Vigan}, {Wahhaj}, \&
  {Wisniewski}}]{boccaletti2015}
{Boccaletti}, A., {Thalmann}, C., {Lagrange}, A.-M., {et~al.} 2015, \nat, 526,
  230

\bibitem[{{Brandt} {et~al.}(2013){Brandt}, {McElwain}, {Turner}, {Abe},
  {Brandner}, {Carson}, {Egner}, {Feldt}, {Golota}, {Goto}, {Grady}, {Guyon},
  {Hashimoto}, {Hayano}, {Hayashi}, {Hayashi}, {Henning}, {Hodapp}, {Ishii},
  {Iye}, {Janson}, {Kandori}, {Knapp}, {Kudo}, {Kusakabe}, {Kuzuhara}, {Kwon},
  {Matsuo}, {Miyama}, {Morino}, {Moro-Mart{\'{\i}}n}, {Nishimura}, {Pyo},
  {Serabyn}, {Suto}, {Suzuki}, {Takami}, {Takato}, {Terada}, {Thalmann},
  {Tomono}, {Watanabe}, {Wisniewski}, {Yamada}, {Takami}, {Usuda}, \&
  {Tamura}}]{brandt2013}
{Brandt}, T.~D., {McElwain}, M.~W., {Turner}, E.~L., {et~al.} 2013, \apj, 764,
  183

\bibitem[{{Currie} {et~al.}(2012){Currie}, {Rodigas}, {Debes}, {Plavchan},
  {Kuchner}, {Jang-Condell}, {Wilner}, {Andrews}, {Kraus}, {Dahm}, \&
  {Robitaille}}]{currie2012}
{Currie}, T., {Rodigas}, T.~J., {Debes}, J., {et~al.} 2012, \apj, 757, 28

\bibitem[{{Debes} {et~al.}(2009){Debes}, {Weinberger}, \&
  {Kuchner}}]{debes2009}
{Debes}, J.~H., {Weinberger}, A.~J., \& {Kuchner}, M.~J. 2009, \apj, 702, 318

\bibitem[{{Donaldson} {et~al.}(2013){Donaldson}, {Lebreton}, {Roberge},
  {Augereau}, \& {Krivov}}]{donaldson2013}
{Donaldson}, J.~K., {Lebreton}, J., {Roberge}, A., {Augereau}, J.-C., \&
  {Krivov}, A.~V. 2013, \apj, 772, 17

\bibitem[{{Dong} {et~al.}(2015){Dong}, {Zhu}, {Rafikov}, \& {Stone}}]{dong2015}
{Dong}, R., {Zhu}, Z., {Rafikov}, R.~R., \& {Stone}, J.~M. 2015, \apjl, 809, L5

\bibitem[{{Esposito} {et~al.}(2014){Esposito}, {Fitzgerald}, {Graham}, \&
  {Kalas}}]{esposito2014}
{Esposito}, T.~M., {Fitzgerald}, M.~P., {Graham}, J.~R., \& {Kalas}, P. 2014,
  \apj, 780, 25

\bibitem[{{Fitzgerald} {et~al.}(2007){Fitzgerald}, {Kalas}, \&
  {Graham}}]{fitzgerald2007}
{Fitzgerald}, M.~P., {Kalas}, P.~G., \& {Graham}, J.~R. 2007, \apj, 670, 557

\bibitem[{{Hashimoto} {et~al.}(2011){Hashimoto}, {Tamura}, {Muto}, {Kudo},
  {Fukagawa}, {Fukue}, {Goto}, {Grady}, {Henning}, {Hodapp}, {Honda},
  {Inutsuka}, {Kokubo}, {Knapp}, {McElwain}, {Momose}, {Ohashi}, {Okamoto},
  {Takami}, {Turner}, {Wisniewski}, {Janson}, {Abe}, {Brandner}, {Carson},
  {Egner}, {Feldt}, {Golota}, {Guyon}, {Hayano}, {Hayashi}, {Hayashi}, {Ishii},
  {Kandori}, {Kusakabe}, {Matsuo}, {Mayama}, {Miyama}, {Morino}, {Moro-Martin},
  {Nishimura}, {Pyo}, {Suto}, {Suzuki}, {Takato}, {Terada}, {Thalmann},
  {Tomono}, {Watanabe}, {Yamada}, {Takami}, \& {Usuda}}]{hashimoto2011}
{Hashimoto}, J., {Tamura}, M., {Muto}, T., {et~al.} 2011, \apjl, 729, L17

\bibitem[{{Hinkley} {et~al.}(2009){Hinkley}, {Oppenheimer}, {Soummer},
  {Brenner}, {Graham}, {Perrin}, {Sivaramakrishnan}, {Lloyd}, {Roberts}, \&
  {Kuhn}}]{hinkley2009}
{Hinkley}, S., {Oppenheimer}, B.~R., {Soummer}, R., {et~al.} 2009, \apj, 701,
  804

\bibitem[{{Janson} {et~al.}(2013){Janson}, {Brandt}, {Kuzuhara}, {Spiegel},
  {Thalmann}, {Currie}, {Bonnefoy}, {Zimmerman}, {Sorahana}, {Kotani},
  {Schlieder}, {Hashimoto}, {Kudo}, {Kusakabe}, {Abe}, {Brandner}, {Carson},
  {Egner}, {Feldt}, {Goto}, {Grady}, {Guyon}, {Hayano}, {Hayashi}, {Hayashi},
  {Henning}, {Hodapp}, {Ishii}, {Iye}, {Kandori}, {Knapp}, {Kwon}, {Matsuo},
  {McElwain}, {Mede}, {Miyama}, {Morino}, {Moro-Mart{\'{\i}}n}, {Nakagawa},
  {Nishimura}, {Pyo}, {Serabyn}, {Suenaga}, {Suto}, {Suzuki}, {Takahashi},
  {Takami}, {Takato}, {Terada}, {Tomono}, {Turner}, {Watanabe}, {Wisniewski},
  {Yamada}, {Takami}, {Usuda}, \& {Tamura}}]{janson2013}
{Janson}, M., {Brandt}, T.~D., {Kuzuhara}, M., {et~al.} 2013, \apjl, 778, L4

\bibitem[{{Janson} {et~al.}(2016){Janson}, {Thalmann}, {Boccaletti}, {Maire},
  {Zurlo}, {Marzari}, {Meyer}, {Carson}, {Augereau}, {Garufi}, {Henning},
  {Desidera}, {Asensio-Torres}, \& {Pohl}}]{janson2016}
{Janson}, M., {Thalmann}, C., {Boccaletti}, A., {et~al.} 2016, \apjl, 816, L1

\bibitem[{{Joos} {et~al.}(2008){Joos}, {Buenzli}, {Schmid}, \&
  {Thalmann}}]{joos2008}
{Joos}, F., {Buenzli}, E., {Schmid}, H.~M., \& {Thalmann}, C. 2008, in Society
  of Photo-Optical Instrumentation Engineers (SPIE) Conference Series, Vol.
  7016, Observatory Operations: Strategies, Processes, and Systems II, 70161I

\bibitem[{{Kalas}(2005)}]{kalas2005}
{Kalas}, P. 2005, \apjl, 635, L169

\bibitem[{{Kalas} {et~al.}(2007){Kalas}, {Duchene}, {Fitzgerald}, \&
  {Graham}}]{kalas2007}
{Kalas}, P., {Duchene}, G., {Fitzgerald}, M.~P., \& {Graham}, J.~R. 2007,
  \apjl, 671, L161

\bibitem[{{Lafreni{\`e}re} {et~al.}(2007){Lafreni{\`e}re}, {Marois}, {Doyon},
  {Nadeau}, \& {Artigau}}]{lafreniere2007}
{Lafreni{\`e}re}, D., {Marois}, C., {Doyon}, R., {Nadeau}, D., \& {Artigau},
  {\'E}. 2007, \apj, 660, 770

\bibitem[{{Maness} {et~al.}(2008){Maness}, {Fitzgerald}, {Paladini}, {Kalas},
  {Duchene}, \& {Graham}}]{maness2008}
{Maness}, H.~L., {Fitzgerald}, M.~P., {Paladini}, R., {et~al.} 2008, \apjl,
  686, L25

\bibitem[{{Marois} {et~al.}(2006){Marois}, {Lafreni{\`e}re}, {Doyon},
  {Macintosh}, \& {Nadeau}}]{marois2006}
{Marois}, C., {Lafreni{\`e}re}, D., {Doyon}, R., {Macintosh}, B., \& {Nadeau},
  D. 2006, \apj, 641, 556

\bibitem[{{Mawet} {et~al.}(2009){Mawet}, {Serabyn}, {Stapelfeldt}, \&
  {Crepp}}]{mawet2009}
{Mawet}, D., {Serabyn}, E., {Stapelfeldt}, K., \& {Crepp}, J. 2009, \apjl, 702,
  L47

\bibitem[{{Meshkat} {et~al.}(2013){Meshkat}, {Bailey}, {Rameau}, {Bonnefoy},
  {Boccaletti}, {Mamajek}, {Kenworthy}, {Chauvin}, {Lagrange}, {Su}, \&
  {Currie}}]{meshkat2013}
{Meshkat}, T., {Bailey}, V., {Rameau}, J., {et~al.} 2013, \apjl, 775, L40

\bibitem[{{Milli} {et~al.}(2012){Milli}, {Mouillet}, {Lagrange}, {Boccaletti},
  {Mawet}, {Chauvin}, \& {Bonnefoy}}]{milli2012}
{Milli}, J., {Mouillet}, D., {Lagrange}, A.-M., {et~al.} 2012, \aap, 545, A111

\bibitem[{{Min} {et~al.}(2012){Min}, {Canovas}, {Mulders}, \&
  {Keller}}]{min2012}
{Min}, M., {Canovas}, H., {Mulders}, G.~D., \& {Keller}, C.~U. 2012, \aap, 537,
  A75

\bibitem[{{Moerchen} {et~al.}(2007){Moerchen}, {Telesco}, {De Buizer},
  {Packham}, \& {Radomski}}]{moerchen2007}
{Moerchen}, M.~M., {Telesco}, C.~M., {De Buizer}, J.~M., {Packham}, C., \&
  {Radomski}, J.~T. 2007, \apjl, 666, L109

\bibitem[{{Muto} {et~al.}(2012){Muto}, {Grady}, {Hashimoto}, {Fukagawa},
  {Hornbeck}, {Sitko}, {Russell}, {Werren}, {Cur{\'e}}, {Currie}, {Ohashi},
  {Okamoto}, {Momose}, {Honda}, {Inutsuka}, {Takeuchi}, {Dong}, {Abe},
  {Brandner}, {Brandt}, {Carson}, {Egner}, {Feldt}, {Fukue}, {Goto}, {Guyon},
  {Hayano}, {Hayashi}, {Hayashi}, {Henning}, {Hodapp}, {Ishii}, {Iye},
  {Janson}, {Kandori}, {Knapp}, {Kudo}, {Kusakabe}, {Kuzuhara}, {Matsuo},
  {Mayama}, {McElwain}, {Miyama}, {Morino}, {Moro-Martin}, {Nishimura}, {Pyo},
  {Serabyn}, {Suto}, {Suzuki}, {Takami}, {Takato}, {Terada}, {Thalmann},
  {Tomono}, {Turner}, {Watanabe}, {Wisniewski}, {Yamada}, {Takami}, {Usuda}, \&
  {Tamura}}]{muto2012}
{Muto}, T., {Grady}, C.~A., {Hashimoto}, J., {et~al.} 2012, \apjl, 748, L22

\bibitem[{{Nesvold} \& {Kuchner}(2015)}]{nesvold2015}
{Nesvold}, E.~R. \& {Kuchner}, M.~J. 2015, \apj, 798, 83

\bibitem[{{Pearce} \& {Wyatt}(2015)}]{pearce2015}
{Pearce}, T.~D. \& {Wyatt}, M.~C. 2015, \mnras, 453, 3329

\bibitem[{{Perrin} {et~al.}(2015){Perrin}, {Duchene}, {Millar-Blanchaer},
  {Fitzgerald}, {Graham}, {Wiktorowicz}, {Kalas}, {Macintosh}, {Bauman},
  {Cardwell}, {Chilcote}, {De Rosa}, {Dillon}, {Doyon}, {Dunn}, {Erikson},
  {Gavel}, {Goodsell}, {Hartung}, {Hibon}, {Ingraham}, {Kerley}, {Konapacky},
  {Larkin}, {Maire}, {Marchis}, {Marois}, {Mittal}, {Morzinski}, {Oppenheimer},
  {Palmer}, {Patience}, {Poyneer}, {Pueyo}, {Rantakyr{\"o}}, {Sadakuni},
  {Saddlemyer}, {Savransky}, {Soummer}, {Sivaramakrishnan}, {Song}, {Thomas},
  {Wallace}, {Wang}, \& {Wolff}}]{perrin2015}
{Perrin}, M.~D., {Duchene}, G., {Millar-Blanchaer}, M., {et~al.} 2015, \apj,
  799, 182

\bibitem[{{Perryman} {et~al.}(1997){Perryman}, {Lindegren}, {Kovalevsky},
  {Hoeg}, {Bastian}, {Bernacca}, {Cr{\'e}z{\'e}}, {Donati}, {Grenon},
  {Grewing}, {van Leeuwen}, {van der Marel}, {Mignard}, {Murray}, {Le Poole},
  {Schrijver}, {Turon}, {Arenou}, {Froeschl{\'e}}, \&
  {Petersen}}]{perryman1997}
{Perryman}, M.~A.~C., {Lindegren}, L., {Kovalevsky}, J., {et~al.} 1997, \aap,
  323

\bibitem[{{Qi} {et~al.}(2015){Qi}, {{\"O}berg}, {Andrews}, {Wilner}, {Bergin},
  {Hughes}, {Hogherheijde}, \& {D'Alessio}}]{qi2015}
{Qi}, C., {{\"O}berg}, K.~I., {Andrews}, S.~M., {et~al.} 2015, \apj, 813, 128

\bibitem[{{Quillen}(2006)}]{quillen2006}
{Quillen}, A.~C. 2006, \mnras, 372, L14

\bibitem[{{Redfield}(2007)}]{redfield2007}
{Redfield}, S. 2007, \apjl, 656, L97

\bibitem[{{Rodigas} {et~al.}(2014){Rodigas}, {Debes}, {Hinz}, {Mamajek},
  {Pecaut}, {Currie}, {Bailey}, {Defrere}, {De Rosa}, {Hill}, {Leisenring},
  {Schneider}, {Skemer}, {Skrutskie}, {Vaitheeswaran}, \&
  {Ward-Duong}}]{rodigas2014}
{Rodigas}, T.~J., {Debes}, J.~H., {Hinz}, P.~M., {et~al.} 2014, \apj, 783, 21

\bibitem[{{Schmid} {et~al.}(2006){Schmid}, {Joos}, \& {Tschan}}]{schmid2006}
{Schmid}, H.~M., {Joos}, F., \& {Tschan}, D. 2006, \aap, 452, 657

\bibitem[{{Schneider} {et~al.}(2014){Schneider}, {Grady}, {Hines}, {Stark},
  {Debes}, {Carson}, {Kuchner}, {Perrin}, {Weinberger}, {Wisniewski},
  {Silverstone}, {Jang-Condell}, {Henning}, {Woodgate}, {Serabyn},
  {Moro-Martin}, {Tamura}, {Hinz}, \& {Rodigas}}]{schneider2014}
{Schneider}, G., {Grady}, C.~A., {Hines}, D.~C., {et~al.} 2014, \aj, 148, 59

\bibitem[{{Schneider} {et~al.}(2005){Schneider}, {Silverstone}, \&
  {Hines}}]{schneider2005}
{Schneider}, G., {Silverstone}, M.~D., \& {Hines}, D.~C. 2005, \apjl, 629, L117

\bibitem[{{Silverstone, M.D.}(2000)}]{silverstone2000}
{Silverstone, M.D.} 2000, PhD thesis, AA (University of California, Los
  Angeles)

\bibitem[{{Soummer} {et~al.}(2012){Soummer}, {Pueyo}, \&
  {Larkin}}]{soummer2012}
{Soummer}, R., {Pueyo}, L., \& {Larkin}, J. 2012, \apjl, 755, L28

\bibitem[{{Tamura} {et~al.}(2006){Tamura}, {Hodapp}, {Takami}, {Abe}, {Suto},
  {Guyon}, {Jacobson}, {Kandori}, {Morino}, {Murakami}, {Stahlberger},
  {Suzuki}, {Tavrov}, {Yamada}, {Nishikawa}, {Ukita}, {Hashimoto}, {Izumiura},
  {Hayashi}, {Nakajima}, \& {Nishimura}}]{tamura2006}
{Tamura}, M., {Hodapp}, K., {Takami}, H., {et~al.} 2006, in Society of
  Photo-Optical Instrumentation Engineers (SPIE) Conference Series, Vol. 6269,
  Society of Photo-Optical Instrumentation Engineers (SPIE) Conference Series,
  62690V

\bibitem[{{Thalmann} {et~al.}(2013){Thalmann}, {Janson}, {Buenzli}, {Brandt},
  {Wisniewski}, {Dominik}, {Carson}, {McElwain}, {Currie}, {Knapp},
  {Moro-Mart{\'{\i}}n}, {Usuda}, {Abe}, {Brandner}, {Egner}, {Feldt}, {Golota},
  {Goto}, {Guyon}, {Hashimoto}, {Hayano}, {Hayashi}, {Hayashi}, {Henning},
  {Hodapp}, {Ishii}, {Iye}, {Kandori}, {Kudo}, {Kusakabe}, {Kuzuhara}, {Kwon},
  {Matsuo}, {Mayama}, {Miyama}, {Morino}, {Nishimura}, {Pyo}, {Serabyn},
  {Suto}, {Suzuki}, {Takami}, {Takato}, {Terada}, {Tomono}, {Turner},
  {Watanabe}, {Yamada}, {Takami}, \& {Tamura}}]{thalmann2013}
{Thalmann}, C., {Janson}, M., {Buenzli}, E., {et~al.} 2013, \apjl, 763, L29

\bibitem[{{Thalmann} {et~al.}(2015){Thalmann}, {Mulders}, {Janson}, {Olofsson},
  {Benisty}, {Avenhaus}, {Quanz}, {Schmid}, {Henning}, {Buenzli}, {M{\'e}nard},
  {Carson}, {Garufi}, {Messina}, {Dominik}, {Leisenring}, {Chauvin}, \&
  {Meyer}}]{thalmann2015}
{Thalmann}, C., {Mulders}, G.~D., {Janson}, M., {et~al.} 2015, \apjl, 808, L41

\bibitem[{{Wyatt}(2008)}]{wyatt2008}
{Wyatt}, M.~C. 2008, \araa, 46, 339

\end{thebibliography}
\end{document}